\documentclass{article}

\usepackage{spconf}

\usepackage{amsmath,amssymb,amsfonts}
\usepackage{algorithmic}
\usepackage{textcomp}
\usepackage{stfloats}
\usepackage{url}
\usepackage{verbatim}
\usepackage{graphicx}
\usepackage{cite}
\usepackage{multirow}
\usepackage{lscape,array}
\usepackage{booktabs}
\usepackage{pifont}
\usepackage{ragged2e}
\usepackage{enumitem}
\usepackage[table]{xcolor}
\usepackage{balance}
\usepackage{hyperref}

\title{Hybrid Pruning: In-Situ Compression of Self-Supervised Speech Models for Speaker Verification and Anti-Spoofing
}
\name{
    \begin{tabular}{c}
        Junyi Peng$^1$, Lin Zhang$^2$, Jiangyu Han$^1$, Old\v{r}ich Plchot$^1$, Johan Rohdin$^1$, \\
        Themos Stafylakis$^{3,4,5}$, Shuai Wang$^6$, Jan \v{C}ernock\'{y}$^1$
    \end{tabular}
}

\address{
    \begin{tabular}{c}
        $^1$Speech@FIT, Brno University of Technology, Czechia 
        $^2$Johns Hopkins University, USA \\
        $^3$Athens University of Economics and Business \quad
        $^4$Omilia \quad
        $^5$Archimedes/Athena R.C., Greece \\
        $^6$Nanjing University, China
    \end{tabular}
}
\begin{document}
\ninept
\maketitle

\begin{abstract}
Although large-scale self-supervised learning (SSL) models like WavLM have achieved state-of-the-art performance in speech processing, their significant size impedes deployment on resource-constrained devices. While structured pruning is a key technique for model compression, existing methods typically separate it from task-specific fine-tuning. This multi-stage approach struggles to create optimal architectures tailored for diverse downstream tasks. In this work, we introduce a unified framework that integrates structured pruning into the downstream fine-tuning process. 
Our framework unifies these steps, jointly optimizing for task performance and model sparsity in a single stage. This allows the model to learn a compressed architecture specifically for the end task, eliminating the need for complex multi-stage pipelines and knowledge distillation. Our pruned models achieve up to a 70\% parameter reduction with negligible performance degradation on large-scale datasets, achieving equal error rates of 0.7\%, 0.8\%, and 1.6\% on Vox1-O, -E, and -H, respectively. Furthermore, our approach demonstrates improved generalization in low-resource scenarios, reducing overfitting and achieving a state-of-the-art 3.7\% EER on ASVspoof5.
\footnote{Models are available at \url{https://huggingface.co/JYP2024/Wedefense_ASV2025_WavLM_Base_Pruning}}
\end{abstract}
\section{Introduction}
\label{sec:intro}

Self-supervised learning has advanced speech processing by providing versatile pretrained models for downstream tasks like automatic speech recognition (ASR), speaker verification (SV), and anti-spoofing \cite{hsu2021hubert, li2023parameter ,wang22_odyssey}. Models like WavLM~\cite{chen2022wavlm}, trained on large-scale speech datasets, achieve strong performance across these tasks. However, their large size (e.g., 94 million parameters for WavLM-Base and over 300 million for WavLM-Large) makes deployment challenging in resource-constrained environments, such as mobile devices or edge systems.

One promising compression approach is structured pruning~\cite{peng23c_interspeech,wang23da_interspeech,han2025efficient,ebrahimipour2025latency}, which removes entire components like attention heads or network layers to yield direct reductions in memory and latency. 
However, current pruning methods are often disconnected from the final application.
They are typically applied separately either during pre-training~\cite{peng23c_interspeech, wang23da_interspeech}, agnostic to the downstream task, or as a post-hoc step after fine-tuning~\cite{han2025efficient, 10095780}. This separation prevents the pruning process from being guided by the specific requirements of the end task, making it difficult to find an optimal compressed architecture. 

To investigate these questions in a controlled yet diverse setting, we focus on SV and anti-spoofing. These tasks provide a compelling testbed due to their reliance on different acoustic cues. SV aims to extract stable, long-term speaker characteristics, while anti-spoofing requires the detection of subtle, transient synthesis artifacts. This contrast makes them an ideal case study for exploring our research questions and testing whether a pruning method can discover truly task-specific architectures.

The key research questions can be summarized as three-fold: 
(1)~\textbf{Flexibility of Representations}: Do we really need to preserve all original layer-wise representations?
Given that final performance depends on the output embeddings, it’s worth investigating whether the model can redistribute representations across layers during pruning without decreasing performance.
(2) \textbf{Task-Dependent Sparsity}: Do distinct downstream tasks like SV and anti-spoofing result in unique architectural pruning patterns? It is unclear whether a single compressed architecture can serve both, or if each requires a task-specific sparsity signature.
(3) \textbf{Influence of Data}: Are pruning patterns driven primarily by dataset characteristics? For example, the optimal sparsity for a model trained on a large corpus like VoxCeleb may differ from one trained on a smaller dataset like CN-Celeb.

\begin{figure*}
    \centering
    \includegraphics[width=0.99\linewidth]{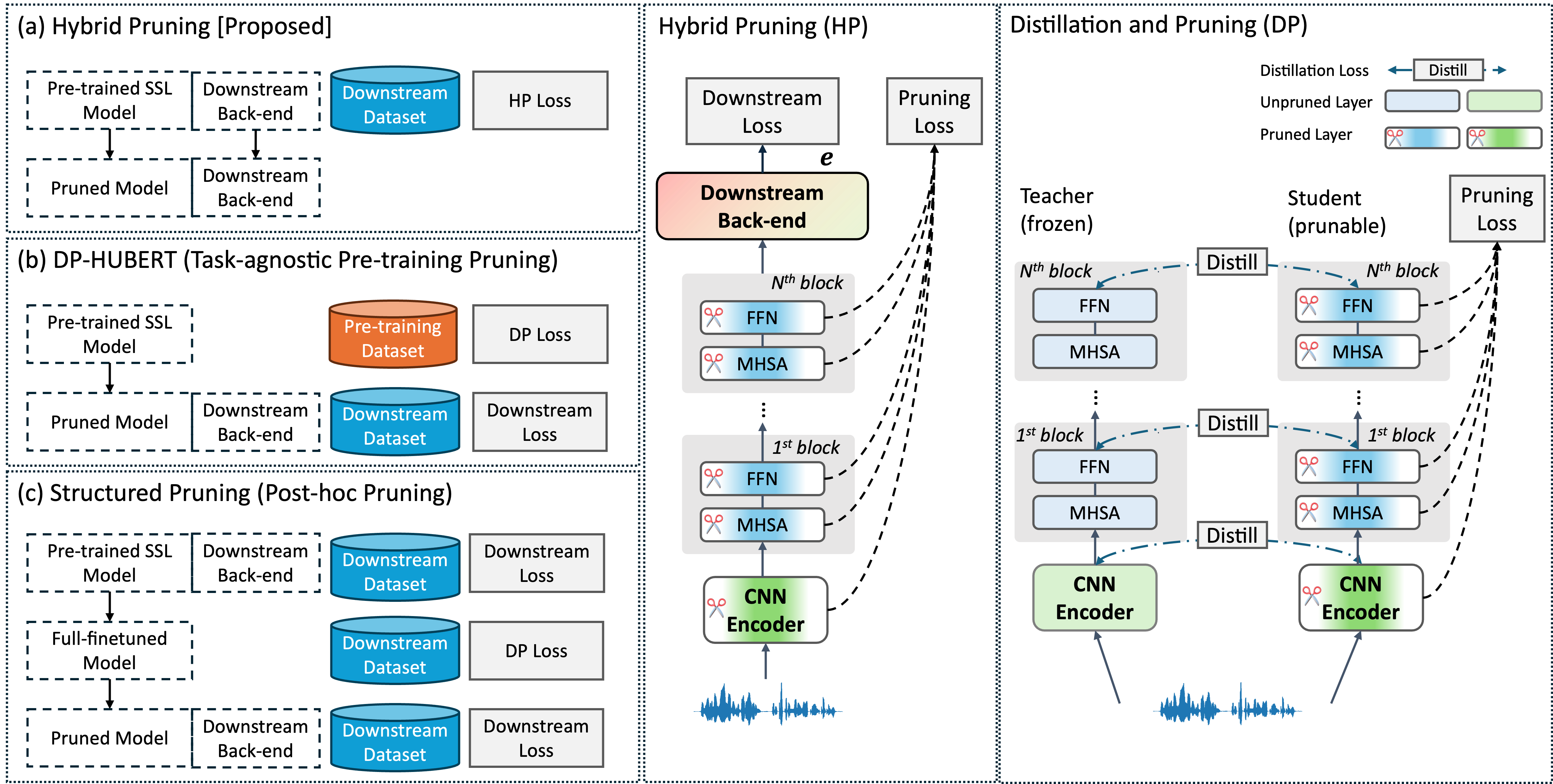}
        \vspace{-0.3cm}
    \caption{The proposed Hybrid Pruning (HP) framework in comparison to prior work. (a) Our single-stage approach jointly optimizes for the downstream task with downstream back-end, such as speaker extractor back-end and structured sparsity, directly learning a pruned architecture. This avoids multi-stage pipelines like (b) task-agnostic pre-training pruning~\cite{peng23c_interspeech} and (c) post-hoc pruning~\cite{han2025efficient}. Notably, our method (center detail) does not require the teacher-student knowledge distillation (right detail) common in other lightweight techniques \cite{liao22_spsc}.}
    \label{fig:enter-label}
        \vspace{-0.5cm}
\end{figure*}

To address these challenges and explore the aforementioned research questions, we propose a hybrid pruning (HP) paradigm that integrates structured pruning and downstream fine-tuning into a unified, efficient training process. 
As illustrated in Figure \ref{fig:enter-label} (a), we integrate learnable pruning masks directly into the SSL model architecture during fine-tuning. The model is then trained with a joint objective that simultaneously minimizes a task-specific loss (e.g., for SV or anti-spoofing) and a pruning loss that encourages sparsity. 
This single-stage approach enables the model to dynamically adapt its architecture by removing non-essential components, guided directly by the downstream objective.

Overall, our contributions are the following: (1) We introduce a unified framework that integrates structured pruning directly into downstream fine-tuning, driven by task-specific classification loss, without the need for separate knowledge distillation or multi-stage pipelines. (2) We conduct a comprehensive evaluation across tasks, including SV (CNCeleb\cite{li2020cn}, VoxCeleb\cite{chung2018voxceleb2}, anti-spoofing (ASVspoof5~\cite{wang24_asvspoof}, SpoofCeleb~\cite{jung2025spoofceleb}), model scales (Base, Large), and dataset sizes (200 hours, 2000 hours). Our method achieves up to 70\% sparsity with minimal performance loss on large datasets and improves results on low-resource datasets by mitigating overfitting. (3) We provide a detailed analysis of the resulting sparsity patterns, revealing task-dependent pruning signatures and showing how data scale influences architectural choices and representational shifts within the pruned model.

\section{Related work}
\textbf{SSL in Speaker and Spoofing Detection.} SSL representations have become foundational in both SV and anti-spoofing tasks. 
However, a widely used pipeline involves fine-tuning the entire large SSL model along with the downstream back-end \cite{peng2022attention, chen2022large, peng2025mhfa, rohdin24_asvspoof} such as ECAPA-TDNN~\cite{desplanques20_interspeech} and MHFA~\cite{peng2022attention}, leading to high computational costs and a risk of overfitting, particularly when downstream data is limited~\cite{peng2023parameter}.
Our work directly addresses this by proposing to simultaneously fine-tune and prune the SSL model. This approach not only simplifies the compression pipeline but also avoids reliance on a potentially suboptimal, pre-finetuned teacher model, allowing the student model to forge its own, more effective optimization path.

\noindent \textbf{Structured Pruning of Speech Models.} Structured pruning is an increasingly popular technique for compressing SSL models. Methods can be broadly categorized as task-agnostic and task-specific approaches. Task-agnostic method, such as DP-HuBERT~\cite{peng23c_interspeech,wang23da_interspeech} as shown in Figure~\ref{fig:enter-label}(b), integrates pruning into the SSL pre-training stage itself. Task-specific, as shown in Figure~\ref{fig:enter-label}(c), apply pruning post-hoc to a model already fine-tuned for a task like ASR, often requiring knowledge distillation to recover performance~\cite{han2025efficient, jiang23d_interspeech,han2025fine}. Many existing downstream-focused methods follow a multi-stage process involving: (1) full fine-tuning, (2) pruning with distillation, and (3) a final recovery step. 
This paper proposes a hybrid approach to address this gap by integrating structured pruning directly into the fine-tuning stage, optimizing the model's architecture and weights jointly in a single pass without dependence on a frozen teacher model.

\section{The Hybrid Pruning Framework}
\label{sec:method}

We introduce Hybrid Pruning (HP), a framework that directly co-optimizes a model's architecture and weights in a single optimization stage. Departing from multi-stage pipelines, HP simultaneously fine-tunes a model for a downstream task while structurally pruning it to a target sparsity. To achieve this with high precision on downstream tasks, we employ a gradient-based $L_0$ regularization method adapted from~\cite{louizos2018learning, wang2020structured}.

\subsection{Dynamic Architectural Adaptation via Stochastic Gates}
\label{sec:dynamic_adaptation}
Our framework begins with a pre-trained SSL model (e.g., WavLM) whose parameters are unfrozen for fine-tuning. We augment this model by inserting learnable \textit{stochastic gates} at each of its prunable structural components, including CNN kernels,  Multi-headed Self-attention (MHSA) heads, and Feed-Forward Network (FFN) neurons. Each gate is a random variable $z_j \in [0, 1]$ whose value determines the contribution of the corresponding component to the network's output. 

To make this gating mechanism differentiable, we employ the Hard Concrete distribution \cite{maddison2016concrete, louizos2017learning} to model each $z_j$. This is a continuous distribution that serves as a relaxation of the discrete Bernoulli distribution but has the critical property of being able to take an exact value of zero with non-zero probability. This allows the network to learn, via gradient descent, to completely ``turn off'' or partially prune certain components. The probability of a gate being zero is controlled by a learnable parameter $\alpha_j$, effectively making the model's architecture itself a subject of the optimization process.

\subsection{Joint Optimization for Performance and Sparsity}
\label{sec:joint_objective}
The core of HP is a joint optimization objective, where we simultaneously minimize a loss with respect to the model and gate parameters while maximizing it with respect to a set of sparsity-controlling parameters. This creates a dynamic equilibrium during training: the model strives to improve its performance, while the Lagrangian multipliers actively `punish' it for being too complex, effectively forcing the model to find the simplest possible architecture that can still perform the task well. This allows for a direct trade-off between task accuracy and model complexity within a single training loop. 
The overall objective is:

\begin{equation}
    \max_{\lambda_{1},\lambda_{2}} \min_{\theta, \alpha} \left( \mathcal{L}_{\text{task}}(\theta) + \mathcal{R}_{\text{prune}}(\theta, \alpha, \lambda_1, \lambda_2) \right)
    \label{eq:final_loss}
\end{equation}
where $\theta$ are the model weights and $\alpha$ are the gate distribution parameters.

\textbf{Task-Specific Loss ($\mathcal{L}_{\text{task}}$):} The first term anchors the model to the downstream objective. For SV, we use the AAM Softmax loss~\cite{deng2019arcface} to learn discriminative embeddings. For anti-spoofing, we employ a standard Binary Cross-Entropy (BCE) loss. This term ensures that any architectural changes made by the pruning process are evaluated based on their impact on final task performance.

\textbf{Pruning Regularizer ($\mathcal{R}_{\text{prune}}$):} The second term drives the model towards a target sparsity level. Given the ideal regularizer is the non-differentiable $L_0$ norm, we adopt the tractable variational approximation from~\cite{louizos2018learning}. To precisely steer the model to a pre-defined target sparsity (i.e. the ratio of pruned parameters to the original model size), $t$, we use an augmented Lagrangian controller \cite{wang2020structured}. The regularizer is therefore the expected value of this controller over the distribution of the stochastic gates:

\begin{equation}
    \mathcal{R}_{\text{prune}} = \mathbb{E}_{q(\theta|\overline{\theta},\alpha)}[\lambda_{1} (||\theta||_{0} - t) + \lambda_{2}(||\theta||_{0} - t)^2]
    \label{eq:prune_regularizer}
\end{equation}
Here, the expectation of the $L_0$ norm can be calculated analytically as a function of the gate parameters $\alpha$. The learnable Lagrange multipliers, $\lambda_1, \lambda_2 \in \mathbb{R}$, are updated via gradient ascent to enforce the sparsity constraint. And $\overline{\theta}$ refers to the deterministic location parameters of the variational distribution $q$.


Upon completion of training, the stochastic gates are finalized to deterministic binary values (0 or 1). All structural components corresponding to a gate value of 0 are \textbf{permanently removed} from the model's computational graph. The result is a physically smaller, faster model that can be deployed for inference without any special libraries or dependencies for sparse computation, thereby achieving a practical reduction in both latency and memory footprint.

\section{Experiments}
\subsection{Setup}\label{sec:exp:setup}

\textbf{Datasets.} 
We conduct a comprehensive evaluation of our proposed framework across two distinct tasks: SV and anti-spoofing.
For SV, we utilize standard large-scale benchmarks. The models are trained on the development set of the VoxCeleb2 dataset\cite{nagrani2017voxceleb, chung2018voxceleb2}. Then the performance is evaluated on the three official trials: Vox1-O, Vox1-E, and Vox1-H. To further assess the generalization capabilities of our method under different language and domain conditions, we also report performance on the CNCeleb corpus~\cite{fan2020cn}.
For anti-spoofing, we train on ASVspoof5~\cite{wang24_asvspoof} and SpoofCeleb~\cite{jung2025spoofceleb}, and test on their respective official sets.

\noindent \textbf{Implementation details.} We utilize two scales of pre-trained SSL models: (1) The \emph{BASE} models, consisting of a CNN encoder and 12 layers of Transformers with approximately 94M parameters. (2) The \emph{Large} models, which include 24 layers of Transformers with approximately 316M parameters. For both SV and anti-spoofing tasks, we employ the same MHFA module with 32 attention heads, which accounts for a total of 1.2 million parameters~\cite{peng2022attention}. During the training of our HP framework, the parameters of this back-end are jointly optimized along with the SSL model's parameters and the pruning masks. For SV tasks, we employ the AAM-softmax loss with a margin of 0.2 and a scaling factor of 32. For the anti-spoofing task, we utilize a standard BCE loss.
To ensure stable convergence during pruning, we implement a warm-up schedule for the target sparsity. The sparsity target is linearly increased from 0 to its final pre-defined value over the first 5 epochs of training.


\begin{table}[t]
\centering
\caption{The EER results in the ASVSpoof5 eval sets varied across different sparsity targets. FLOPs are calculated based on a 4-second audio input.}
\label{tab:my-table1}
\scalebox{0.82}{
\begin{tabular}{llcccc}
\toprule
\multirow{2}{*}{Sparsity} & \multirow{2}{*}{Methods} & \multirow{2}{*}{\#Param} & \multirow{2}{*}{FLOPs} & \multicolumn{2}{c}{Eval} \\
\cmidrule(lr){5-6} 
& & & & EER (\%) & minDCF \\
\midrule
\multicolumn{4}{c}{WavLM-SLIM (Best single system in ASVspoof5)~\cite{zhu2024learn}} & 5.16 & 0.149\\
\midrule
\rowcolor{gray!15} 0\% & WavLM Base & 95.6 M & 57.4 G & 4.56 & 0.116 \\
\midrule
\multirow{3}{*}{10\%} & DP-HUBERT & 86.2 M & 48.0 G & 5.13 & 0.139\\
& Structured Pruning & 86.7 M & 51.7 G & 5.57 & 0.154\\
& Ours & 86.0 M & 51.9 G & \textbf{3.75} & \textbf{0.103} \\
\midrule
\multirow{3}{*}{30\%} & DP-HUBERT & 67.3 M & 36.4 G & 7.23 & 0.200\\
& Structured Pruning & 67.3 M & 36.7 G & 5.62 & 0.149\\
& Ours & 67.3 M & 39.1 G & \textbf{5.14} & \textbf{0.143} \\
\midrule
\multirow{3}{*}{50\%} & DP-HUBERT & 48.5 M & 25.8 G & 11.73 & 0.321 \\
& Structured Pruning & 48.4 M & 25.2 G & 10.22 & 0.269 \\
& Ours & 48.4 M & 26.9 G & \textbf{8.74} & \textbf{0.233} \\
\bottomrule
\end{tabular}
}
\end{table}

\begin{table}[t]
\centering
\caption{Performance and efficiency trade-offs on the VoxCeleb dataset. EERs\% are reported for the VoxCeleb1-O/E/H test sets.  Speedups measured on an AMD EPYC 7A53 CPU and an AMD MI250 GPU are reported relative to the unpruned model.}
\label{tab:my-table2}
\scalebox{0.85}{
\begin{tabular}{lccccccc}
\toprule
\multirow{2}{*}{Model} & \multirow{2}{*}{Sparsity} & \multirow{2}{*}{\#Param} & \multicolumn{2}{c}{Speedup} & \multicolumn{3}{c}{VoxCeleb}  \\
\cmidrule(lr){4-5} \cmidrule(lr){6-8}
& & & CPU & GPU & O & E & H \\
\midrule
\rowcolor{gray!15} \multirow{4}{*}{WavLM Base+} & 0\%   & 95.6\,M & - & - & 0.70 & 0.69 & 1.40 \\
& 60\% &  38.9 M & 2.2$\times$ & 2.0$\times$ & 0.70 & 0.78 & 1.50  \\
& 70\% &  29.5 M & 2.9$\times$ & 2.6$\times$ & 0.73 & 0.84 & 1.61  \\
& 80\% &  19.9 M & 3.8$\times$ & 3.4$\times$ & 0.92 & 1.02 & 1.91  \\
\bottomrule
\end{tabular}
}
\end{table}

\begin{figure}[t]
    \centering
    \includegraphics[width=0.99\linewidth]{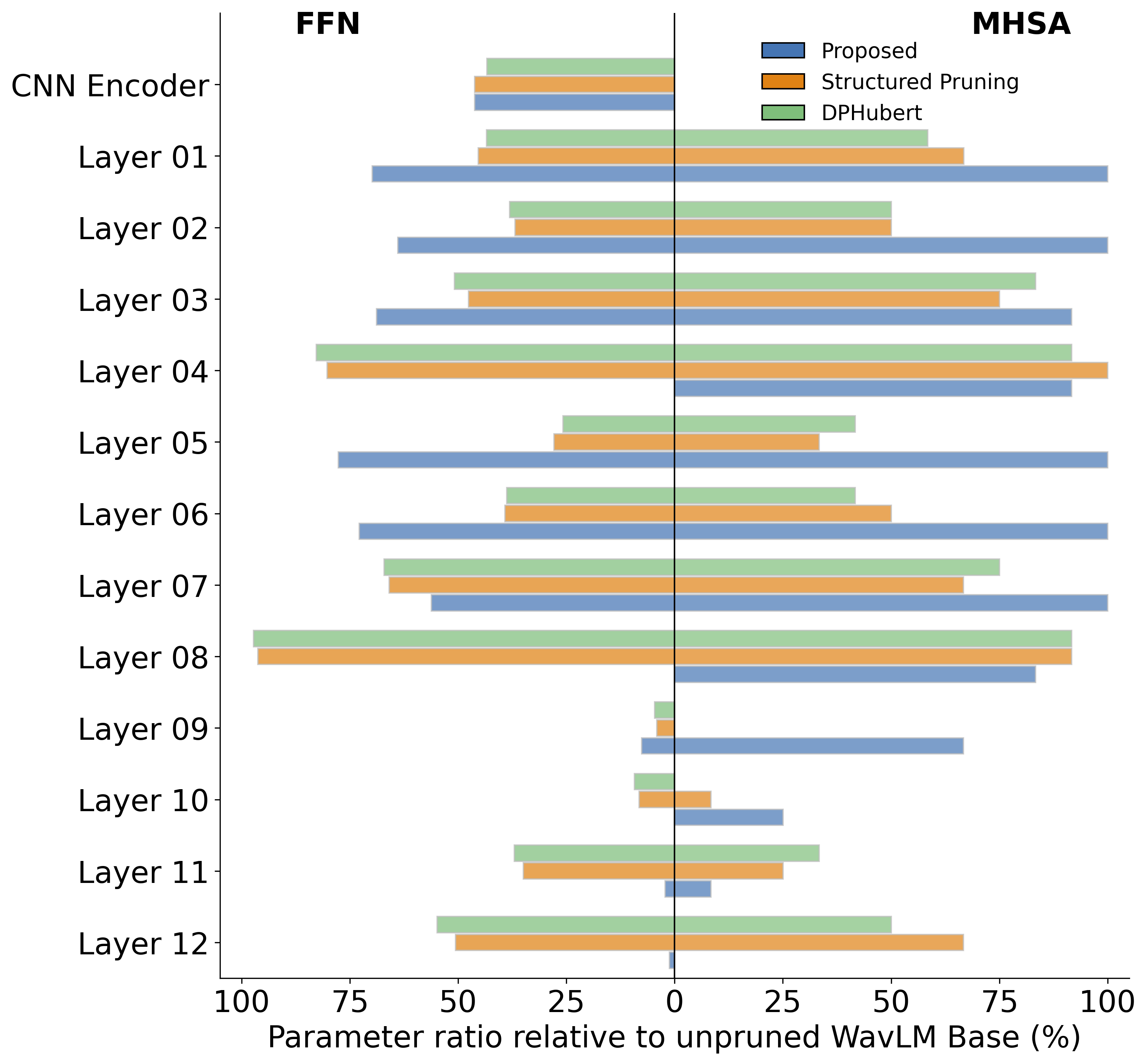}
        \vspace{-0.3cm}
    \caption{Layer-wise pruning patterns (on FFN and MHSA) for WavLM Base across different methods. All methods were trained on the ASVspoof5 dataset and pruned to 50\% target sparsity.}
    \label{fig:exp1}
        \vspace{-0.2cm}
\end{figure}

\begin{figure}[tb]
    \centering
    \includegraphics[width=0.99\linewidth]{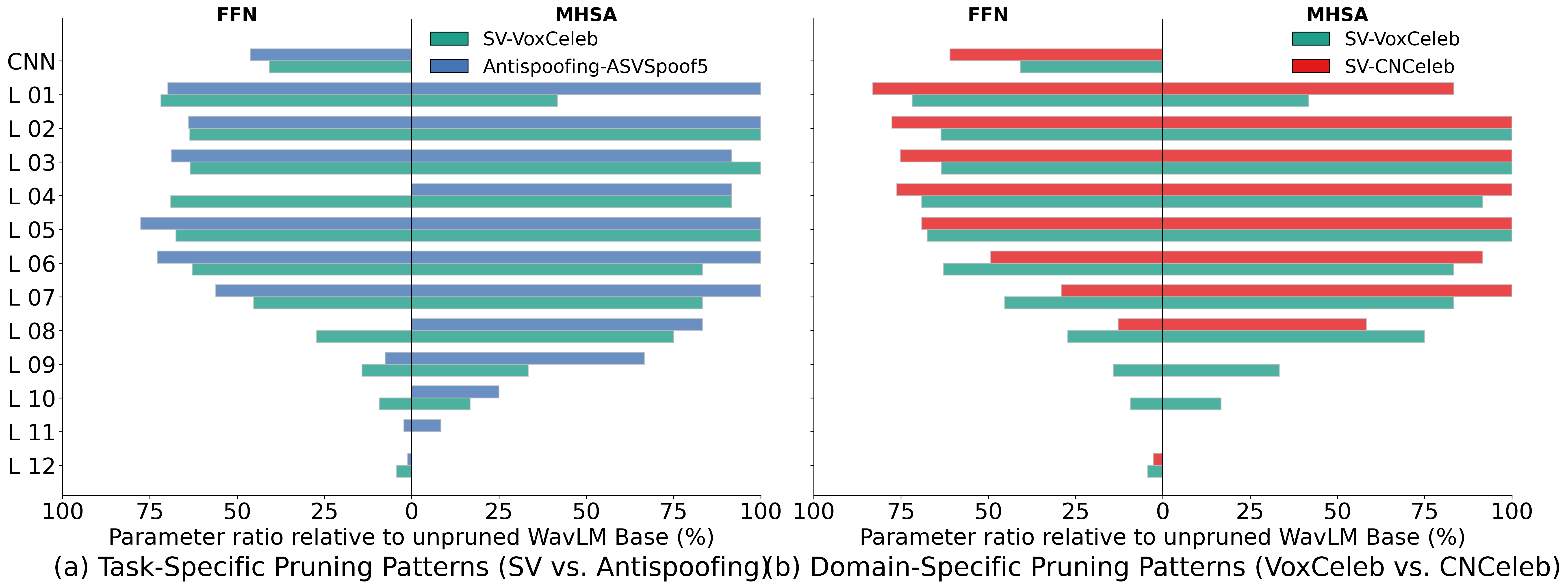}
        \vspace{-0.5cm}
    \caption{Layer-wise pruning patterns, comparing (a) task-specificity between SV and anti-spoofing, and (b) domain-specificity for SV between VoxCeleb and CNCeleb datasets. }
    \label{fig:exp2}    
    \vspace{-0.3cm}
\end{figure}

\subsection{Main Results on SV and Anti-Spoofing Benchmarks}

We evaluate our HP framework on the anti-spoofing and SV tasks.

For anti-spoofing, we use the standard ASVspoof5 benchmark, comparing our HP against two strong pruning baselines: a task-agnostic method (DP-HUBERT) and a post-hoc approach (Structured Pruning). As shown in Table \ref{tab:my-table1}, our HP method consistently and significantly outperforms both baselines at every sparsity level. 
Additionally, our method is the only one that improves performance over the unpruned baseline at 10\% sparsity with 3.70\% EER, which shows our method acts not only as a compression tool but as a powerful regularizer, as further analyzed in Sec \ref{sec:4-4}. Furthermore, the FLOPs of our pruned model differ from the baselines at the same sparsity, which indicates distinct sub-architectures are being learned.

As shown in Table~\ref{tab:my-table2}, HP attains efficient \emph{inference} while preserving accuracy for SV tasks. At 60\% sparsity, WavLM Base+ reduces parameters from 95.6M to 38.9M and yields $2.2\times$ CPU and $2.0\times$ GPU \emph{inference speedup} over the dense WavLM, with EERs of 0.70\%, 0.78\%, and 1.50\% on Vox1-O/E/H, comparable to the dense model (0.70\%, 0.69\%, 1.40\%). 


\begin{figure}[t]
    \centering
    \includegraphics[width=0.95\linewidth]{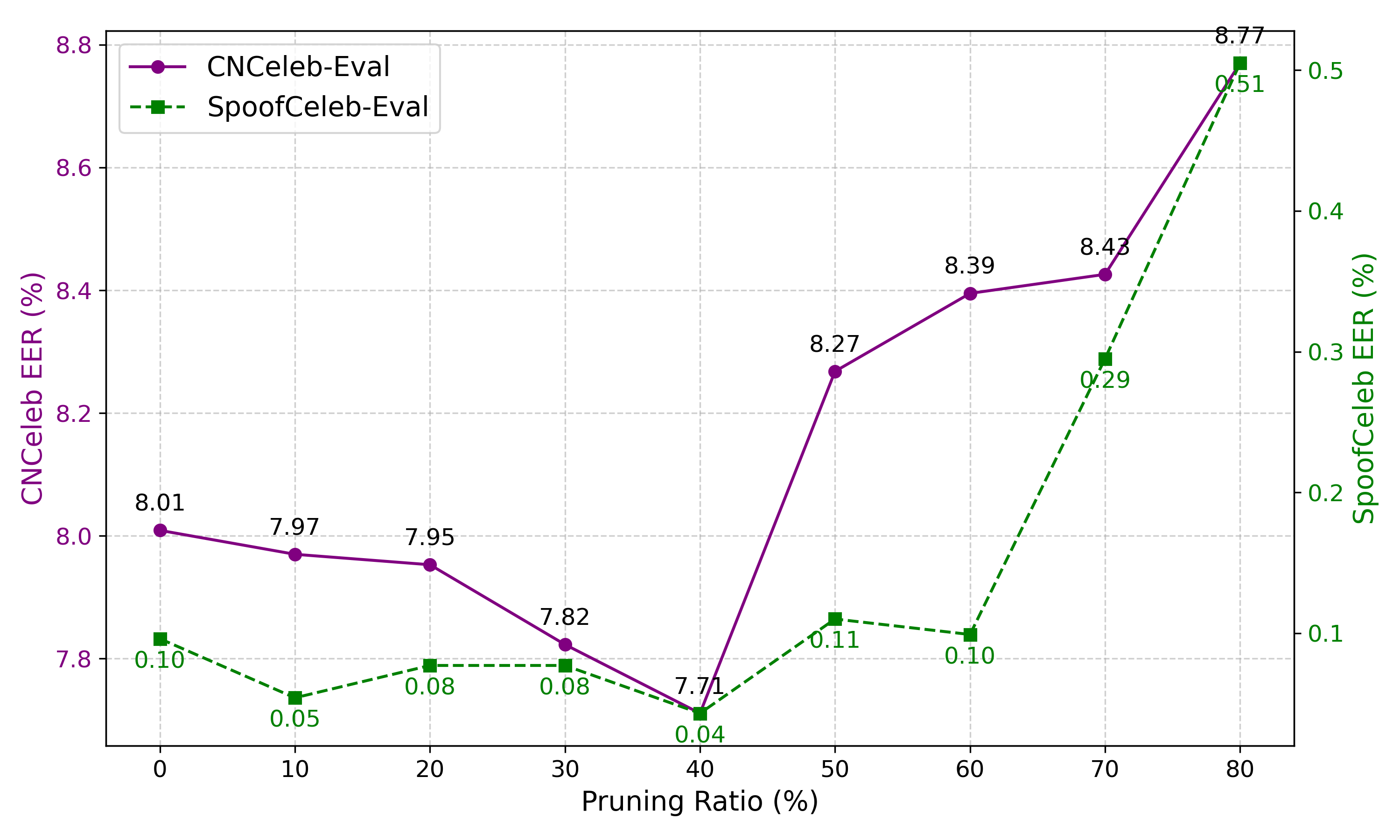}
    \vspace{-0.3cm}
    \caption{EER performance of the HP framework on the CNCeleb-Eval (purple) and SpoofCeleb-Eval (green) datasets.}
    \label{fig:enter-label3}
        \vspace{-0.3cm}
\end{figure}

\subsection{
Task- and Domain-Adaptive Architectures after Pruning}

Our HP framework learns non-uniform, specialized sub-architectures that adapt to both the downstream task and the data domain with 50\% sparsity, in contrast to the more uniform strategies of the baselines (Figure \ref{fig:exp1}). This task-specificity is evident in Figure \ref{fig:exp2}(a), where the pruning patterns for SV and anti-spoofing are nearly opposite: the anti-spoofing model preserves lower-layer MHSA modules to detect acoustic artifacts, while the SV model retains upper-middle layers to capture speaker identity. Moreover, the framework shows domain-specificity (Figure \ref{fig:exp2}b); the SV model trained on the more diverse CNCeleb dataset prunes upper layers more aggressively than when trained on VoxCeleb. 

\subsection{
Generalization Gains from Moderate Pruning\\
}
\label{sec:4-4}
A key finding is that moderate pruning can act as an effective regularizer, improving generalization of over-parameterized SSL-based systems compared to unpruned ones. As illustrated in Figure~\ref{fig:enter-label3}, both the results on CNCeleb and SpoofCeleb exhibit a distinct ``U-shaped'' performance curve, where an optimal level of sparsity outperforms the baseline. This suggests that standard fine-tuning of SSL models may lead to suboptimal generalization, possibly due to overfitting to the downstream set. By removing redundant parameters, our HP method reduces model capacity in a controlled way, helping the model focus on learning more robust and discriminative features.

\begin{figure}[t]
    \centering
    \includegraphics[width=0.9\linewidth]{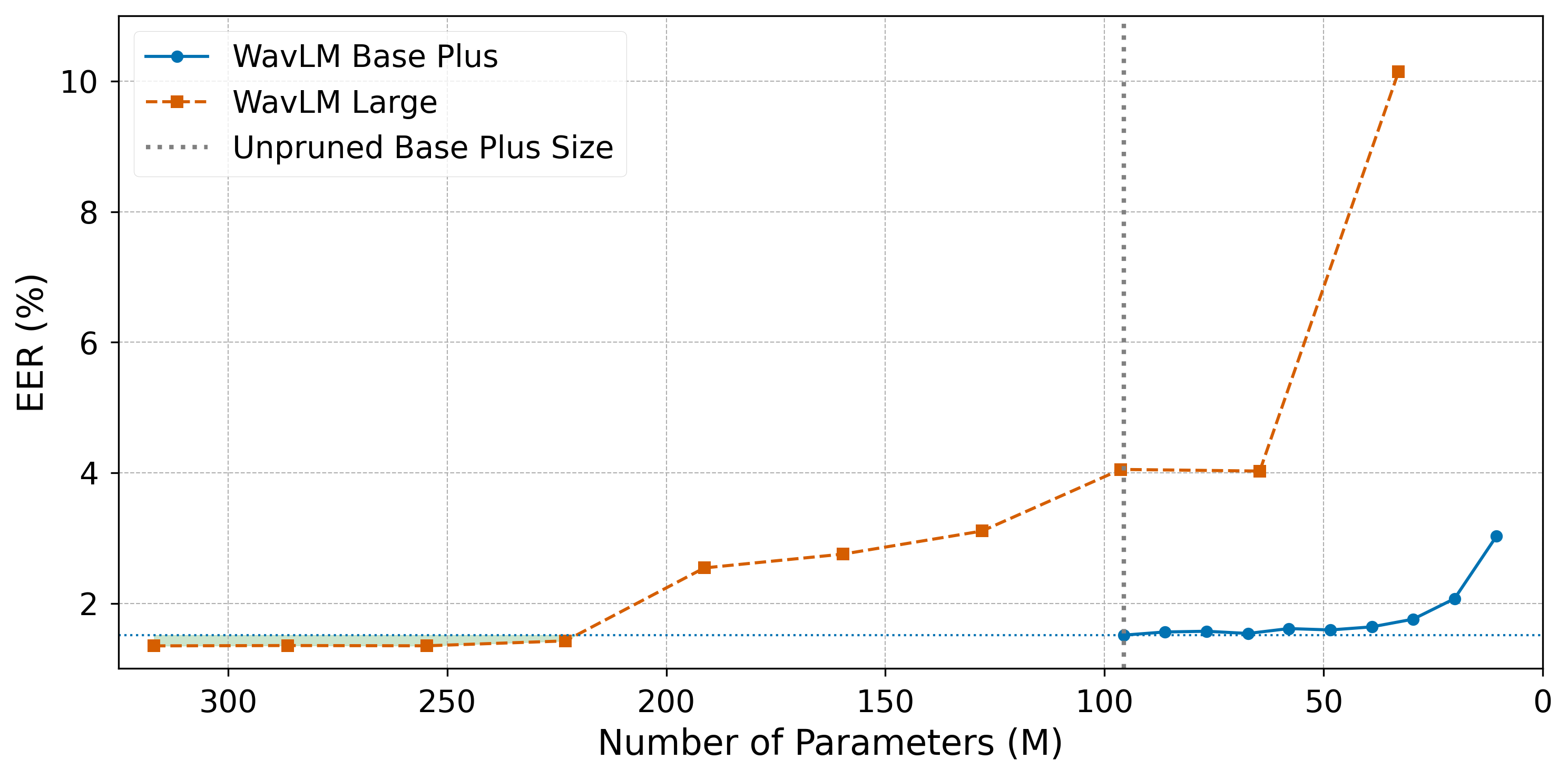}
        \vspace{-0.3cm}
    \caption{EER vs. number of parameters for pruned WavLM Base Plus and WavLM Large models on the Vox1-H test sets.}
    \label{fig:enter-label4}
    \vspace{-0.3cm}
\end{figure}

\subsection{
Pruning Effectiveness Across Model Scales
}
To investigate the impact of model scale on pruning effectiveness, we conducted experiments on both the WavLM Base Plus and Large models, with the results shown in Figure \ref{fig:enter-label4}. The pruned Base model consistently outperforms the pruned Large model, suggesting that starting with a larger model is not always the optimal strategy. Moreover, our HP framework can prune the Base model by over 70\% with negligible performance degradation compared to its unpruned version. It is noted that selecting a suitable foundation model and applying fine-grained pruning is a more effective path to an efficient system than simply compressing the largest available model.

\section{Conclusion}
In this work, we introduced Hybrid Pruning (HP), a single-stage framework that integrates structured pruning into downstream fine-tuning to efficiently compress large SSL models. Our experiments on SV and anti-spoofing benchmarks demonstrate that HP can achieve up to a 70\% parameter reduction with negligible performance degradation. Notably, we found that HP acts as a regularizer, improving generalization in low-resource scenarios by mitigating overfitting. Further analysis revealed that the framework learns task-specific architectures, retaining different layers for SV versus anti-spoofing, and that pruning a smaller base model can be more effective than compressing a larger one. Ultimately, HP provides a principled and efficient path for deploying high-performance SSL models on resource-constrained devices, with future work potentially extending to other tasks like speech large language models.


{\footnotesize
\bibliographystyle{IEEEtran}
\bibliography{mybib}
}
\end{document}